\documentclass[pra,preprintnumbers]{revtex4}
\usepackage{graphicx}

\begin{document}
\newcommand{\bl}{{\bf l}}
\newcommand{\bq}{{\bf q}}
\newcommand{\bk}{{\bf k}}
\newcommand{\bkp}{{\bf k_\perp}}
\newcommand{\bkz}{{\bf k_z}}
\newcommand{\bV}{{\bf V}}
\newcommand{\ba}{{\bf a}}
\newcommand{\bXp}{{\bf X}^\prime}
\newcommand{\bX}{{\bf X}}
\newcommand{\bJ}{{\bf J}}
\newcommand{\bR}{{\bf R}}
\newcommand{\bF}{{\bf F}}
\newcommand{\bZ}{{\bf Z}}
\newcommand{\bp}{{\bf p}}
\newcommand{\bP}{{\bf P}}
\newcommand{\bPp}{{\bf P}^\prime}
\newcommand{\bA}{{\bf A}}
\newcommand{\br}{{\bf r}}
\newcommand{\bu}{{\bf u}}
\newcommand{\be}{{\bf \epsilon}}
\newcommand{\eh}{\hat{\epsilon}}
\newcommand{\bb}{{\bf b}}
\newcommand{\rmi}{\rm i}
\newcommand{\rme}{\rm e}
\newcommand{\rmd}{{\rm d}}
\newcommand{\rmv}{{\rm v}}
\newcommand{\rmR}{{\rm R}}
\newcommand{\rmr}{{\rm r}}
\newcommand{\rms}{{\rm s}}
\newcommand{\rmx}{{\rm x}}
\newcommand{\rmt}{{\rm t}}
\newcommand{\rma}{{\rm a}}
\newcommand{\rmb}{{\rm b}}
\newcommand{\w}{\omega}
\newcommand{\la}{\lambda}
\newcommand{\g}{\sf g}
\newcommand{\gb}{\bar{\sf g}}
\newcommand{\p}{\bf p}
\newcommand{\lab}{\bar{\lambda}}
\newcommand{\e}{\epsilon}
\newcommand{\psib}{\bar{\psi}}
\newcommand{\psip}{\psi^{\prime}}
\newcommand{\psibp}{{\psi^{\prime}}}
\newcommand{\z}{z}
\newcommand{\zb}{\bar{z}}
\newcommand{\zp}{z^{\prime}}
\newcommand{\zbp}{\bar{z^{\prime}}}
\newcommand{\K}{K}
\newcommand{\Kb}{\bar{K}}
\newcommand{\Kp}{K^{\prime}}
\newcommand{\Kbp}{\bar{K^{\prime}}}
\newcommand{\f}{\rm f}
\newcommand{\fb}{\bar{\rm f}}
\newcommand{\fp}{\rm f^{\prime}}
\newcommand{\fbp}{\bar{\rm f^{\prime}}}
\newcommand{\h}{\rm h}
\newcommand{\hb}{\bar{\rm h}}
\newcommand{\hp}{\rm h^{\prime}}
\newcommand{\hbp}{\bar{\rm h^{\prime}}}

\title{Moving Atom-Field Interaction: Correction to Casimir-Polder Effect from Coherent Back-action}
\author{S. Shresta}
\email{sanjiv@physics.umd.edu} \affiliation{Department of Physics, University of Maryland, College Park, Maryland 20742-4111}
\author{B. L. Hu}
\email{hub@physics.umd.edu} \affiliation{Department of Physics, University of Maryland, College Park, Maryland 20742-4111}
\author{Nicholas G. Phillips}
\email{Nicholas.G.Phillips@gsfc.nasa.gov} \affiliation{ SSAI, Laboratory for Astronomy and Solar Physics, Code 685, NASA/GSFC, Greenbelt, Maryland 20771}
\preprint{umdpp03-014}
\date{\today}
\begin{abstract}
The Casimir-Polder force is an attractive force between a polarizable atom and a conducting or dielectric boundary. Its original computation was in terms of the Lamb shift of the atomic ground state in an electromagnetic field (EMF) modified by boundary conditions along the wall and assuming a stationary atom. We calculate the corrections to this force due to a moving atom, demanding maximal preservation of entanglement generated by the moving atom-conducting wall system. We do this by using non-perturbative path integral techniques which allow for coherent back-action and thus can treat non-Markovian processes. We recompute the atom-wall force for a conducting boundary by allowing the bare atom-EMF ground state to evolve (or self-dress) into the interacting ground state. We find a clear distinction between the cases of stationary and adiabatic motions. Our result for the retardation correction for adiabatic motion is up to twice as much as that computed for stationary atoms. We give physical interpretations of both the stationary and adiabatic atom-wall forces in terms of alteration of the virtual photon cloud surrounding the atom by the wall and the Doppler effect.
\end{abstract}
\maketitle

\section{Introduction}
The physical system studied in this paper is an atom in a polarizable ground state near a conducting wall. The interaction of the atom with the quantum electromagnetic field (EMF) vacuum, whose spatial modes are restricted by the wall with imposed boundary conditions, generates a force that pulls it toward the conducting wall (for general discussion see Ref.~\cite{Milonni}). The details of such a force is important in any experiments and applications in which an atom is held near a surface by a trapping scheme using evanescent waves or magnetic fields. The atom-wall force is divisible into two parts. First, there is the electrostatic attraction that the atom feels toward its image on the other side of the wall, called the van~der~Waals (vdW) force. Second is a quantum mechanical modification of the vdW force first calculated by Casimir and Polder~\cite{caspol}. They dubbed the quantum modification "retardation" of the vdW force, because its source is the non-instantaneous transverse EMF. Extensions of Casimir and Polder's results for a polarizable atom were later derived by many authors~\cite{barut87,ford02}, including for an atom in a cavity~\cite{barton1} and near a dielectric wall~\cite{wylie84,wueberlein}. Closest in philosophy to what is done in this paper is the work of Milonni in Ref.~\cite{Milonni82}. There, the author computes the second order alteration of the EMF mode functions due to the presence of the atom, from which the ground state energy shift is the expectation value of the interaction Hamiltonian in the altered vacuum \footnote{The author of Ref.~\cite{Milonni82} refers to this method as radiation reaction. We would advise against using this terminology because it is different from the usual meaning referring radiation reaction to the force exerted on a charged object due to its emitted radiation, which manifests as a classical effect.}. However, the author neglects time dependence in the mode functions and thus neglects effects due to Doppler shifts of the EMF modes. Recently, retardation correction of the vdW force has been demonstrated experimentally~\cite{suk1,lan1}. Verification of the Casimir-Polder force can be viewed as a demonstration of the entangled quantum behavior of the entire system, since it involves the dressing of the atom by the EMF vacuum.

Although Casimir and Polder and others' calculations do treat the quantum entanglement in the system, analysis up to now has been restricted to stationary atoms. It has been assumed (wrongly, as we shall show) that such a method can also treat the adiabatic motion of the atom. Adiabatic motion means in this context that as the atom moves, it continuously shifts into the position dependent stationary dressed ground state on a timescale much shorter than the timescale of motion. Treatments assuming that the atom is stationary or is instantaneously static exclude correlations that are developed in the system during the motion. The key point is that the adiabatic and stationary dressed vacuum states are not the same. An example where this situation is encountered generically and dealt with in depth is in cosmology, specifically, quantum field processes in an expanding universe~\cite{BirrellDavies}. For stationary systems a vacuum state is well defined at all times (due to the existence of a Killing vector), but not for arbitrary dynamics, especially fast motion. However, for slow dynamics, adiabatic vacuum states can be defined and renormalization procedures constructed~\cite{ParkerFulling74,Hu74,FullingParkerHu74}. The adiabatic method we use here is similar in spirit (though not in substance, as our purpose is somewhat different from that in cosmology). To predict motional effects, entanglement in the evolution needs to be accounted for theoretically. We use the influence functional (IF) method here, which keeps track of full coherence in the evolution to derive the force between the atom and the wall while allowing the atom to move adiabatically. In the case of a stationary atom, our result is in exact agreement with the Casimir-Polder force. In the case of an adiabatically moving atom, we find a coherent retardation correction up to twice the stationary value, thus our coherent QED calculation will make verifiable predictions. This paper shows the derivation and explains the cause due to coherent back-action. Section II outlines the model and details of the calculation. The results for stationary and adiabatic motion are then given in Section III, and discussed in Section IV.

\section{Model and Approach}
In contrast to obtaining the force via the gradient of the ground state energy shift, we obtain it through the expectation value of an atom's center of mass (COM) momentum. Our system consists of an atom placed near a conducting wall. We assume an initially factorized state of the atom in its ground state and the EMF in its vacuum. A path integral technique is used to derive the ground state-EMF vacuum transition amplitude of the evolving system. Inclusion of coherent back-action allows the system to self-dress~\cite{selfdressing,selfdressing3} and preserves maximal entanglement in the non-Markovian evolution of an atom-EMF quantum system. The expectation value of the momentum operator is then computed. In the path integral, Grassmannian and bosonic coherent states are used to label the atomic and EMF degrees of freedom, respectively. The position and momentum basis are used for the atom's center of mass degree of freedom. The major approximation applied here is a second order vertex approximation. With the second order vertex, the propagator is partially resummed to all orders of the coupling constant. The result is a non-perturbative propagator which yields coherent long time dynamics~\cite{ctan, CPP}. The mass of the atom and the size of its external wavepacket are kept finite throughout the calculation. Only at the end of the calculation do we allow the mass of the atom to go to infinity and its extension shrunk to a point, while retaining finite terms due to their effect on the dynamics.

Highlights of the calculation are given in this section and details are given in the Appendices. In Section IIA the Hamiltonian and spatial mode functions that describe the system are introduced. In Section IIB the transition amplitude of the EMF vacuum with the atom in its ground state is calculated in a coherent state path integral, with an effective action expanded to second order in the coupling (equivalent to a second order vertex resummation), and semiclassically in the COM motion. The momentum expectation value and the retardation correction force is then calculated from the transition amplitude in Section IIC.

\subsection{The Hamiltonian}
The spinless non-relativistic QED Hamiltonian is given by \footnote{We would like to point out that we are using the minimal coupling form of the non-relativistic QED Hamiltonian, not the multipolar form. The two forms are related by a gauge transformation, with the '${\bf p}\cdot{\bf A}$' and '${\bf A}^2$' interaction terms appearing in the minimal coupling form and the '${\bf d}\cdot{\bf E}$' and R\"ontgen interaction terms appearing in the multipolar form~\cite{multipolar1}. A common confusion between the two forms comes about because it is the \emph{total} minimal coupling form and the \emph{total} multipolar form which are equivalent within a gauge transformation. If any interaction terms are dropped in either form of the Hamiltonian (e.g. the $\bA^2$ or R\"ontgen terms), then the equivalence is broken. Dropping interaction terms can also cause erroneous results in the computation of physical quantities~\cite{rontgen1,rontgen3}. Desiring to avoid this pitfall, we have been careful to use the \emph{total} minimal coupling form, without dropping any interaction terms. There is, however, one issue with our use of the minimal coupling form. Only in the multipolar form are the internal and external degrees of freedom exactly separable. That is especially important in bound systems of charged particles in which the masses of the particles are commensurate. However, since the bound system in our study is an atom, for which the nuclear mass dominates over the mass of the electrons, the atomic and electronic degrees of freedom are well separated and the use of the minimal coupling Hamiltonian is justified.}
\begin{eqnarray}\label{hamiltonian0}
H =\frac{{\bf P}^2}{2M} +\frac{1}{2m}(\bp -\rme\bA)^2 +\rme V(\bX) +H_b.
\end{eqnarray}
The first term is the COM kinetic energy of an atom with mass $M$. The second term is the kinetic energy of the electron sitting in the transverse EMF. The third term is the potential energy of the electron around the atomic nucleus. The last term is the energy of the free EMF. After taking the dipole approximation, and restricting to two internal levels of the atom, the Hamiltonian in minimal coupling form becomes (see Appendix A of Ref.~\cite{ABL} without the rotating wave approximation)
\begin{eqnarray}
\label{hamiltonian1} H =\frac{{\bf P}^2}{2M} +\hbar\w_0 S_+ S_- +\hbar\sum_\bk \w_\bk \rmb_\bk^\dagger \rmb_\bk +H_{I1} +H_{I2} = H_0 + H_I.
\end{eqnarray}
The operators $S_{\pm}$ are the up and down operators of the atomic qubit and $\w_0$ is the atomic transition frequency. The operators $\rmb_\bk$ and $\rmb_\bk^\dagger$ are the EMF mode annihilation and creation operators, and $\w_\bk$ are the frequencies of the EMF modes. The two parts of the interaction Hamiltonian are
\begin{eqnarray}
\label{interaction1} H_{I1} &=& \hbar\sum_{\bk e} \frac{\g}{\sqrt{\w_\bk}} [\bp_{eg} S_+ +\bp_{ge} S_-]\cdot[\bu_\bk \rmb_\bk +\bu_\bk^\dagger \rmb_\bk^\dagger] \\
\label{interaction2} H_{I2} &=& \hbar\sum_{\bk\bl} \frac{\la^2}{\sqrt{\w_\bk \w_\bl}} [\bu_\bk\cdot\bu_\bl \rmb_\bk \rmb_\bl +\bu_\bk^\dagger\cdot\bu_\bl(\delta_{\bk\bl} +2\rmb_\bk^\dagger \rmb_\bl ) +\bu_\bk^\dagger\cdot\bu_\bl^\dagger \rmb_\bk^\dagger \rmb_\bl^\dagger].
\end{eqnarray}
The vector $\bp_{eg}$ is the dipole transition matrix element, which is defined as $\bp_{eg} = \langle e| \bp | g\rangle = -\rmi m\w_0 \langle e| \br | g\rangle$. The vectors $\bu_\bk$ contain the photon polarization vectors $\hat{\be}_\bk$ and the spatial mode functions $f_\bk(\bX)$, i.e., $\bu_\bk(\bX) = \hat{\be}_\bk f_\bk(\bX)$. The coupling constants are $\g = -\sqrt{\frac{8\pi^2\alpha c}{m^2}}$ and $\la=\sqrt{\frac{4\pi^2\hbar\alpha c}{m}}$, with $\alpha$ being the fine structure constant.

In the presence of a conducting plane the spatial mode functions of the EMF which satisfy the imposed boundary conditions are the TE and TM polarization modes~\cite{Milonni82},
\begin{eqnarray}
\label{mode functions 1} \bu_{\bk 1}(\bX) &=& \sqrt{\frac{2}{L^3}}\hat{\bk}_\|\times\hat{\bZ} \sin(k_Z Z) e^{\rmi\bk_\|\cdot\bX} \\
\label{mode functions 2}\bu_{\bk 2}(\bX) &=& \sqrt{\frac{2}{L^3}}\frac{1}{k}[k_\|\hat{\bZ} \cos(k_Z Z) -\rmi k_Z \hat{\bk}_\| \sin(k_Z Z)] e^{\rmi\bk_\|\cdot\bX},
\end{eqnarray}
and their complex conjugates.

\subsection{The Transition Amplitude}
The transition amplitude between the initial and final coherent states with initial and final positions is given by
\begin{equation}
\langle \bX_f, \{\zb_{\bk f}\}, \psib_f ;\rmt+\tau | \exp[-\frac{\rmi}{\hbar}\int_{\rmt}^{\rmt+\tau} H(\rms) \rmd\rms] | \bX_i, \{\z_{\bk i}\}, \psi_i ;\rmt \rangle.
\end{equation}
The transition amplitude relevant to the atom-wall force is the amplitude that the atom moves from $\bX_i$ to $\bX_f$ without the emission of any physical photons.  This is a very good assumption, since the probability for {\it physical} photon emission is extremely small~\cite{selfdressing3}. The initial and final states are thus characterized by the atom being in its ground state and the EMF in vacuum, with arbitrary COM position states. The initial and final coherent state labels can be set to zero to reflect those states, although during the evolution the system evolves freely, and the motion of the COM is affected by recoil from emission and re-absorption of virtual photons,
\begin{equation}
K[\bX_f; \rmt+\tau, \bX_i; \rmt] = \langle \bX_f; \rmt+\tau | \exp[-\frac{\rmi}{\hbar}\int_{\rmt}^{\rmt+\tau} H(\rms) \rmd\rms] | \bX_i; \rmt \rangle.
\end{equation}
Normally, a variational approach would be a sensible way to compute the functional integrals that make up the transition amplitude. However, since in this case both the anti-resonant as well as resonant rotating wave terms are included in the Hamiltonian (i.e., no RWA), the variational equations for the Grassmann variables will have bosonic sources even when the EMF is taken to be in the vacuum. We know from earlier work that when a Grassmann field variable has a bosonic source, the variational technique cannot unambiguously define the evolution of the Grassmann variable. A better way is to leave the transition amplitude as a discrete product of infinitesmal propagators. The necessary functional integrals can then be computed recursively. Details are in Appendix A. After the EMF and Grassmann path integrals are evaluated, the transition amplitude from the initial motional state ${\bf X}_i$ to the final motional state ${\bf X}_f$ (while keeping the same initial and final atomic ground state and EMF vacuum) is given to $O(\rme^2)$ vertex by
\begin{eqnarray}
\label{general transition amplitude functional} K[\bX_f; \rmt+\tau, \bX_i; \rmt] = \int D\bX \exp\bigg\{ \rmi\int_{\rmt}^{\rmt+\tau} \bigg[\frac{M\dot{\bX}^2}{2\hbar} &+&\rmi\bp_z^2 \int_{\rmt}^\rms \rmd\rmr \sum_\bk \frac{\g^2}{\w_\bk} \mbox{ }\rme^{-\rmi(\w_\bk +\w_0)(\rms-\rmr)}\bu_\bk (\bX(\rms))\cdot\bu_\bk^*(\bX(\rmr)) \nonumber\\
&-&\sum_\bk \frac{\la^2}{\w_\bk} \bu_\bk^*(\bX(\rms))\cdot\bu_\bk(\bX(\rms))  +O(\rme^4) \bigg]\rmd\rms \bigg\}
\end{eqnarray}
where $\bp_z^2 = \langle {\rm g}| \bp_z^2 | {\rm g} \rangle$ is the ground state expectation value of $\bp_z^2$.

A semi-classical approximation to the transition amplitude Eq.~(\ref{general transition amplitude functional}) is obtained by evaluating the action along its classical path. This will neglect the fluctuation terms of order $O(\frac{1}{M})$. The classical path is the straight line path plus terms of order $O(\frac{\rme^2}{M})$,
\begin{equation}
\bX_c(\rms) =\bX_i +\frac{\bX_f -\bX_i}{\tau}(\rms-\rmt) +O\bigg(\frac{\rme^2}{M}\bigg) =\bX_c^0 (\rms) +O\bigg(\frac{\rme^2}{M}\bigg).
\end{equation}
Evaluating the transition amplitude along that path gives
\begin{eqnarray}
\label{classical transition amplitude} K[\bX_f; \rmt+\tau, \bX_i; \rmt] = \bigg( \frac{M}{2\pi\rmi\hbar\tau}\bigg )^{3/2} \exp\bigg\{ \rmi\int_{\rmt}^{\rmt+\tau} \bigg[\frac{M\dot{\bX_c^0}^2}{2\hbar} &+&\rmi\bp_z^2 \int_{\rmt}^\rms \rmd\rmr \sum_\bk \frac{\g^2}{\w_\bk} \mbox{ }\rme^{-\rmi(\w_\bk +\w_0)(\rms-\rmr)}\bu_\bk (\bX_c^0(\rms))\cdot\bu_\bk^*(\bX_c^0(\rmr)) \nonumber\\
&-&\sum_\bk \frac{\la^2}{\w_\bk} \bu_\bk^*(\bX_c^0(\rms)) \cdot\bu_\bk(\bX_c^0(\rms))  +O(\rme^4/M)\bigg]\rmd\rms \bigg\}.
\end{eqnarray}
Using the spatial mode functions of Eqs.~(\ref{mode functions 1}-\ref{mode functions 2}) in the above gives the semi-classical transition amplitude in the presence of a conducting wall (see Eq.~(\ref{classical wall amplitude})).

\subsection{momentum expectation and force}
Given the above expression for the transition amplitude and an initial center of mass wavefunction for the atom, $\Psi(\bP)$, the momentum expectation and the force on the atom (the time derivative of the expectation momentum) can be computed. The momentum expectation is
\begin{equation}
\label{momentum expectation} \langle \hat{\bP} \rangle (\rmt+\tau) = \frac{\hbar}{N} \int \frac{\rmd\bP_f}{(2\pi)^3} \mbox{ }\bP_f\int\rmd\bX_i \rmd\bXp_i \mbox{ }K[\bP_f;\rmt+\tau|\bX_i; \rmt] \mbox{ }\Psi(\bX_i) \Psi^*(\bX^\prime_i) \mbox{ }K^*[\bP_f;\rmt+\tau|\bXp_i; \rmt],
\end{equation}
with the normalization factor
\begin{equation}
N =  \int \frac{\rmd\bP_f}{(2\pi)^3} \int\rmd\bX_i \rmd\bXp_i \mbox{ }K[\bP_f;\rmt+\tau|\bX_i; \rmt] \mbox{ }\Psi(\bX_i) \Psi^*(\bX^\prime_i) \mbox{ }K^*[\bP_f;\rmt+\tau|\bXp_i; \rmt].
\end{equation}
The initial wavefunction can be taken to be a Gaussian centered at $(\bR, \bP_0)$ with the standard deviations $(\sigma, 1/\sigma)$. Such a choice will allow for the possibility that the atom and the wall are moving toward or away from one another. Following the line of calculation detailed in Appendix B, a momentum moment generating function is computed in the limits $M\rightarrow\infty$ and $\sigma\rightarrow 0$ such that $\frac{\bP_0}{M}\rightarrow\bV$ and $\sigma^2 M\rightarrow\infty$ (see Eq.~(\ref{generating function})). From the generating function the momentum expectation value can be computed,
\begin{equation}
\bP(\rmt+\tau) = \frac{\hbar}{\rmi Z(0)} \frac{\partial Z(\bJ)}{\partial\bJ} \Bigg|_{\bJ=0}.
\end{equation}
In the above limits
\begin{eqnarray}
\label{general momentum} \bP(\rmt+\tau) = \bP_0 &-&\frac{2\rmi\la^2\hbar}{L^3} \sum_\bk \frac{\bk_z \cos^2\theta}{\w_\bk} \int_{t}^{\rmt+\tau} \rmd\rms \mbox{ }\rme^{-2\rmi\bk_z \cdot(\bR +\bV (\rms-\rmt))} \nonumber\\
&+&\frac{\g^2 \bp_z^2 \hbar}{L^3} \sum_\bk \frac{\bk_z \cos^2\theta}{\w_\bk} \int_{\rmt}^{\rmt+\tau} \rmd\rms \int_{\rmt}^{s} \rmd\rmr \mbox{ }\rme^{-\rmi\bk_z \cdot(2\bR +\bV (\rms+\rmr-2\rmt))} \bigg[ \rme^{-\rmi(\w_k +\w_0)(\rms-\rmr)} -\rme^{\rmi(\w_k +\w_0)(\rms-\rmr)} \bigg].
\end{eqnarray}
The momentum depends on the position and velocity only through the distance from the wall and the velocity toward or away from the wall, so motions parallel to the wall have no effects. Define $\rmR=\hat{\bf e}_z \cdot\bR$ and $\rmv=\hat{\bf e}_z \cdot\bV$, with $\hat{\bf e}_z$ defined as positive away from the wall. Taking the time derivative of the momentum expectation value will give the force that is exerted on the atom by the transverse EMF in the presence of the wall. Doing so, as well as applying the Thomas-Reiche-Kuhn sum rule,
\begin{eqnarray}
\la^2 = \frac{\g^2 \bp_z^2}{\omega_0},
\end{eqnarray}
and rewriting in terms of the static ground state polarizability, $\alpha_0$, the force is
\begin{eqnarray}
\label{general force} \bF_c(\rmR,\rmv,\rmt+\tau) = &-&\frac{2\pi\rmi\alpha_0 \hbar\w_0^2}{L^3} \sum_\bk \frac{\bk_z \cos^2\theta}{\w_\bk} \mbox{ }\rme^{-2\rmi\bk_z \cdot(\bR +\bV\tau)} \nonumber\\
&+&\frac{\pi\alpha_0 \hbar\w_0^3}{L^3} \sum_\bk \frac{\bk_z \cos^2\theta}{\w_\bk} \int_{\rmt}^{\rmt+\tau} \rmd\rms \mbox{ }\rme^{-\rmi\bk_z \cdot(2\bR +\bV (\tau+\rms-\rmt))} \bigg[ \rme^{-\rmi(\w_k +\w_0)(\rmt+\tau-\rms)} -\rme^{\rmi(\w_k +\w_0)(\rmt+\tau-\rms)} \bigg].
\end{eqnarray}
The subscript "c" is a reminder that the force calculated from the transverse field is the retardation correction to the electrostatic force. Inspection of the force reveals that it is a sum over recoil momenta weighted by amplitudes which depend on the distance of the atom from the wall and the velocity of the atom. As will be discussed in Section IV, the recoil momenta come from virtual photon emission and re-absorption. In that sense the net force reflects an interference phenomenon, since it is the net sum of many different possible virtual processes.

\section{Results}
\subsection{Stationary Atom}
If the atom is stationary, then setting $\rmv =0$ gives the retardation force to be
\begin{eqnarray}
\label{stationary correction force} \bF_c^{(0)}(\rmR,\rmv=0,\rmt+\tau) = &-&\frac{2\pi\rmi\alpha_0 \hbar\w_0^2}{L^3} \sum_\bk \frac{\bk_z \cos^2\theta}{\w_\bk} \mbox{ }\rme^{-2\rmi\bk_z \cdot\bR} \nonumber\\
&+&\frac{\pi\alpha_0 \hbar\w_0^3}{L^3} \sum_\bk \frac{\bk_z \cos^2\theta}{\w_\bk} \int_{\rmt}^{\rmt+\tau} \rmd\rms \mbox{ }\rme^{-2\rmi\bk_z \cdot\bR} \bigg[ \rme^{-\rmi(\w_k +\w_0)(\rmt+\tau-\rms)} -\rme^{\rmi(\w_k +\w_0)(\rmt+\tau-\rms)} \bigg].
\end{eqnarray}
Combining the correction force with the electrostatic force gives the total force on a stationary atom,
\begin{eqnarray}
\bF_{sa}(\rmR,\rmt+\tau) = -\hat{\bf e}_z \frac{3\alpha_0 \hbar\w_0}{8 \rmR^4} +\bF_c^{(0)}(\rmR,\rmv=0,\rmt+\tau).
\end{eqnarray}
The stationary atom force exhibits a transient behavior when the atom first "sees" itself in the wall. Then, on a timescale of several atom-wall round trip light travel times it asymptotes to a constant steady state value. The transient behavior is plotted in Fig.~(1) and  Fig.~(2) for an optical transition frequency in an alkali atom.
\begin{figure}
\includegraphics{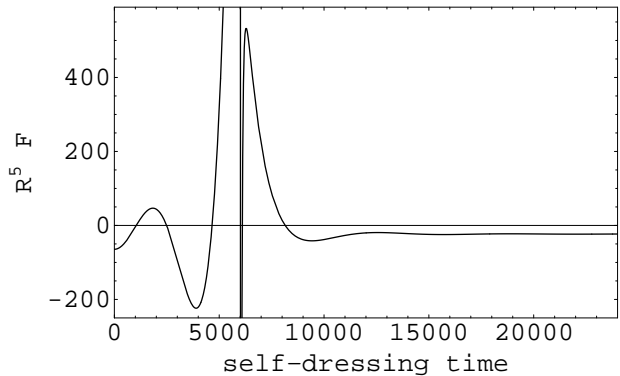}
\caption{This plot shows the value of the atom-wall force at $\rmR=3000$ vs time in atomic units. The spike at $\tau=6000$ is the time at which a photon emitted at $\tau=0$ will have just returned. Before $\tau=6000$ the force is experiencing transient behavior, and afterward it rings down to the stationary atom value.}
\end{figure}
\begin{figure}
\includegraphics{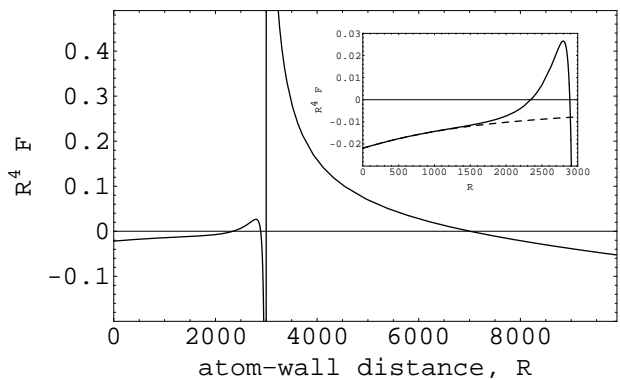}
\caption{This plot shows a snapshot of the coefficient of the $\frac{1}{\rmR^4}$ behavior of the the atom-wall force at a time $\tau=6000$ in atomic units. The location of the spike at $\rmR=3000$ corresponds to the location at which a photon emitted at $\tau=0$ will have just returned to $\rmR=3000$. At locations $\rmR<3000$ the force has begun to asymptote to its steady state behavior, and those at $\rmR>3000$ are still experiencing transient behavior. The inset image is a magnification near the wall. The dotted line is the coefficient of the $\frac{1}{\rmR^4}$ dependence of a stationary atom.
}
\end{figure}
The steady state value of the stationary atom-wall force can also
be determined analytically to be
\begin{eqnarray}
\bF_{sa}(\rmR, \tau>>2\rmR/c) = -\hat{\bf e}_z \frac{3\alpha_o\hbar\w_0}{8\rmR^4} -\hat{\bf e}_z \frac{\alpha_o\hbar\w_0^2}{4\pi } \bigg(\frac{\rmd}{\rmd\rmR}\bigg)^3 \int_0^\infty \frac{\rmd k}{k c+\w_0} \frac{\sin(2k\rmR)}{2k\rmR},
\end{eqnarray}
which can be simplified to
\begin{eqnarray}
\label{stationary force}\bF_{sa}(\rmR, \tau>>2\rmR/c) = \hat{\bf e}_z \frac{\alpha_o\hbar\w_0^2}{8\pi } \bigg(\frac{\rmd}{\rmd \rmR}\bigg)^3 \frac{1}{\rmR} \int_0^\infty \frac{\rmd\rmx}{\rmx^2 +\w_0^2} \mbox{ }\rme^{-2\rmR \rmx/c}.
\end{eqnarray}
From Eq.~(\ref{stationary force}) the potential which a stationary atom feels is easily found to be
\begin{eqnarray}
\label{stationary potential}U_{sa}(\rmR) = -\frac{\alpha_o\hbar\w_0^2}{8\pi } \bigg(\frac{\rmd}{\rmd \rmR}\bigg)^2 \frac{1}{\rmR} \int_0^\infty \frac{\rmd\rmx}{\rmx^2 +\w_0^2} \mbox{ }\rme^{-2\rmR \rmx/c},
\end{eqnarray}
with asymptotic limits
\begin{equation} \label{stationary limits}
\begin{array}{lr}
U_{sa}(\rmR) \rightarrow -\frac{\alpha_o\hbar\w_0}{8 } \frac{1}{\rmR^3} & \mbox{ for } \rmR << \frac{c}{\w_0} \\
U_{sa}(\rmR) \rightarrow -\frac{3\alpha_o\hbar c}{8\pi } \frac{1}{\rmR^4} & \mbox{ for } \rmR >> \frac{c}{\w_0}
\end{array},
\end{equation}
which exactly reproduces the results of energy gradient approaches. Although the results are the same as those previously derived, the interpretation behind how the results are obtained is different. The energy gradient approach can be described as a kinematic approach since the atom-EMF system is assumed to be held static in its entangled dressed ground state. The self-dressing approach used here, on the other hand, allows the atom-EMF entanglement to evolve dynamically. That is, the atom and EMF system, beginning in a factorized state, evolves into a stationary dressed state (i.e. it self-dresses). When the atom is stationary the two forces match because after some time to 'get acquainted', the self-dressing atom does indeed evolve into the stationary dressed state. It should be stressed that the agreement between the results of the two methods demonstrates the coherence of the self-dressing method as applied here.

\subsection{Adiabatic Motion}
We now show that the prediction given by the self-dressing method of the retardation correction force for a slowly moving atom differs from the energy gradient prediction \footnote{Strictly speaking, such a comparison can not be made since energy gradient approaches implicitly assume the atom to be stationary, although they are often assumed to be applicable to moving atoms often with no justification.}. The key difference is that as a moving atom and the EMF get acquainted, they evolve into an entangled dressed state which is different from the stationary atom dressed state. The reason for the difference is the Doppler shift of the EMF modes in combination with the presence of the wall. We will discuss this point in more detail in Section IV.

\subsubsection{adiabatic evaluation}
The retardation force for a moving atom can be determined from Eq.~(\ref{general force}) by applying a separation of short time scale dynamics from long time scale dynamics, and how they affect each other. The adiabaticity condition is applied here in the same way that it is applied in standard methods for determining the dipole force on an atom in a laser beam~\cite{leshouches}. There, assuming that the atom's position is constant on short timescales, the optical Bloch equations are solved for the steady state values of the internal state density matrix elements. On long time-scales the matrix elements are replaced by their steady state values and put into the Heisenberg equation of motion for the atomic COM momentum. Such a procedure is justified when the internal and external dynamics evolve on vastly different timescales. The analogous separation here will be of the short timescale describing the self-dressing of the atom-EMF system and the long timescale describing the motion of the atom.

In order to be explicit about the timescale separation let us first rewrite Eq.~(\ref{general force}) with the definition $\rmx=\rms-\rmt$, and remember that $\rmt$ is the time at which the atom-EMF system begins to evolve from a factorized state,
\begin{eqnarray}
\label{general force rewrite} \bF_c(\rmR_\rmt,\rmv_\rmt,\tau) = &-&\frac{2\pi\rmi\alpha_0 \hbar\w_0^2}{L^3} \sum_\bk \frac{\bk_z \cos^2\theta}{\w_\bk} \mbox{ }\rme^{-2\rmi\bk_z \cdot(\bR_\rmt +\bV_\rmt \tau)} \nonumber\\
&+&\frac{\pi\alpha_0 \hbar\w_0^3}{L^3} \sum_\bk \frac{\bk_z \cos^2\theta}{\w_\bk} \int_{0}^\tau \rmd\rmx \mbox{ }\rme^{-\rmi\bk_z \cdot(2\bR_\rmt +\bV_\rmt (\tau +\rmx))} \bigg[ \rme^{-\rmi(\w_k +\w_0)(\tau -\rmx)} -\rme^{\rmi(\w_k +\w_0)(\tau -\rmx)} \bigg],
\end{eqnarray}
so that the short timescale dynamics (parameterized by $\tau$ and $\rmx$) is explicitly separated from the long timescale dynamics (parameterized by $\rmt$) on which $\bR_\rmt$ and $\bV_\rmt$ evolve. An adiabatic evaluation of the retardation correction for a moving atom can be extracted from a Taylor series expansion of Eq.~(\ref{general force rewrite}),
\begin{eqnarray}\label{force expansion}
\bF_c(\rmR_\rmt,\rmv_\rmt,\tau) &=& \sum_{n=0}^\infty \frac{\rmv_\rmt^n}{n!} \mbox{ }\bF^{(n)}(\rmR_\rmt,\rmv_\rmt=0,\tau)
\end{eqnarray}
where $n$ denotes the $n$th derivative with respect to velocity. The Taylor series expansion is an equivalent representation of the LHS as long as the RHS converges. Each function $\bF^{(n)}(\rmR_\rmt,\rmv_\rmt=0,\tau)$ exhibits a transient behavior while the atom first "sees" itself in the wall (during times $\tau \sim \frac{2\rmR}{c}$) and asymptotes to steady state behavior on a timescale of several round trip light travel times. The adiabatic approximation is applied at this point by replacing each function $\bF^{(n)}(\rmR_\rmt,\rmv_\rmt=0,\tau)$ by its asymptotic behavior
\begin{eqnarray}
\label{adiabatic momentum nth order} \bF^{(n)}(\rmR_\rmt,\rmv_\rmt=0,\tau) \rightarrow \bF^{(n)}_{ss}(\rmR_\rmt,\tau) &=& -\hat{\bf e}_z \frac{\tau^n}{2^n} \frac{\alpha_0 \hbar\w_0^2}{4\pi} \bigg(\frac{\rmd}{\rmd \rmR_\rmt}\bigg)^{n+3} \int_0^\infty \frac{\rmd k}{k c+\w_0} \frac{\sin(2k\rmR_\rmt)}{2k\rmR_\rmt},
\end{eqnarray}
which means replacing the Taylor expansion, Eq.~(\ref{force expansion}), by its steady state form,
\begin{eqnarray}\label{ss force expansion}
\bF_c(\rmR_\rmt,\rmv_\rmt,\tau) \rightarrow \bF_c^{ss}(\rmR_\rmt,\rmv_\rmt,\tau) &=& \sum_{n=0}^\infty \frac{\rmv_\rmt^n}{n!} \mbox{ }\bF^{(n)}_{ss}(\rmR_\rmt,\tau)
\end{eqnarray}
This step is analogous to replacing the internal state density matrix by its steady state value in adiabatic computations of the dipole force on an atom in a laser beam. Replacing the Taylor expansion by its steady state behavior is adiabatic because it assumes that the expansion terms asymptote to their dressed state form on a timescale much shorter than the timescale on which either the position or velocity of the atom changes. More specifically, for the change in position, the adiabatic condition means that during a round trip light travel time the atom-wall distance has very little relative change, $\rmv \frac{2\rmR}{c} << \rmR$, which is equivalent to the condition that the atomic velocity be non-relativistic,
\begin{eqnarray}
\frac{\rmv}{c} << \frac{1}{2} \mbox{ }.
\end{eqnarray}
Similarly, the adiabatic condition for the change in velocity is that it has very little relative change during a round trip light travel time, $\frac{\bF_{net}}{M} \frac{2\rmR}{c} << \bV$, which can be restated as the net force not changing the kinetic energy of the atom much during a light travel time,
\begin{eqnarray}
(F_{net}\mbox{ }\rmv) \frac{\rmR}{c} << \frac{1}{2} M\rmv^2,
\end{eqnarray}
since $F_{net}\mbox{ }\rmv$ is the power that the net force puts into the atoms mechanical motion. Both conditions are satisfied in typical experimental setups.

Note that rather than tending to a constant steady state value, the terms in the Taylor expansion, Eq.~(\ref{adiabatic momentum nth order}), asymptote to steady state polynomial time dependence, the source of the polynomial time dependence being the $\bk_z\cdot\bV$ Doppler shift term in the exponents of Eq.~(\ref{general force rewrite}). In distinction to the stationary atom case those polynomial time dependencies will lead to non-zero \emph{partial} time derivatives as well as the convective changes due simply to motion the of the atom
\begin{eqnarray}
\frac{\rmd}{\rmd\rms}\mbox{ }\bF_c = \bigg( \frac{\rmd\rmR}{\rmd\rms} \frac{\partial}{\partial\rmR} + \frac{\rmd\rmv}{\rmd\rms}\frac{\partial}{\partial\rmv} +\frac{\partial}{\partial\rms} \bigg) \bF_c.
\end{eqnarray}
The differential change in $\bF_c$ can then be split into two parts, one coming from the convective change and the other from the partial time derivative,
\begin{eqnarray}
\rmd\bF_c = \rmd\bF_c\big|_{convective} +\rmd\rms\mbox{ }\frac{\partial\bF_c}{\partial\rms}.
\end{eqnarray}
The convective differential change is the differential change in the force not including any short timescale time dependence, in other words, the steady state expression at $\tau=0$,
\begin{eqnarray}
\rmd\bF_c = \rmd\bF_c^{ss}\bigg|_{\tau=0} +\rmd\rms\mbox{ }\frac{\partial\bF_c}{\partial\rms},
\end{eqnarray}
with, from Eq.~(\ref{ss force expansion}),~$\rmd\bF_c^{ss}\big|_{\tau=0} = \rmd\bF_c^{(0)}$.  The behavior of the force on long time-scales is computed by integrating the differential change from an initial time at which $\rmv=0$ up to the final time,
\begin{eqnarray}\label{zeroth integrated partial}
\bF_c(\rmt) = \bF_c^{(0)}(\rmt) +\int^{\rmt}_{\rmt_o}\rmd\rms_1\mbox{ }\frac{\partial\bF_c}{\partial\rms}(\rms_1),
\end{eqnarray}
where it has been substituted that $\bF_c(\rmt_0)=\bF_c^{(0)}(\rmt_0)$ (since $\rmv=0$ at $\rmt_0$).  A similar analysis for the differential of the first partial time derivative gives,
\begin{eqnarray}
\rmd\bigg(\frac{\partial\bF_c}{\partial\rms}\bigg) = \rmd\bigg(\frac{\partial\bF_c^{ss}}{\partial\rms}\bigg)_{\tau=0} +\rmd\rms\mbox{ }\frac{\partial}{\partial\rms}\bigg(\frac{\partial\bF_c}{\partial\rms}\bigg)
\end{eqnarray}
from which,
\begin{eqnarray}
\frac{\partial\bF_c}{\partial\rms}(\rms_1) = \frac{\rmv_{\rms_1}}{2} \frac{\rmd}{\rmd\rmR} \bF_c^{(0)}(\rms_1) +\int^{\rms_1}_{\rmt_o}\rmd\rms_2\mbox{ }\frac{\partial^2\bF_c}{\partial\rms^2}(\rms_2).
\end{eqnarray}
Carrying on similar analysis (and rewriting in terms of the zeroth order expansion term) leads to the general expression
\begin{eqnarray}\label{general integrated partials}
\frac{\partial^n\bF_c}{\partial\rms^n}(\rms_n) = \frac{\rmv_{\rms_1}\rmv_{\rms_2}..\rmv_{\rms_n}}{2^n} \frac{\rmd^n}{\rmd\rmR^n} \bF_c^{(0)}(\rms_n) +\int^{\rms_n}_{\rmt_o}\rmd\rms_{n+1}\mbox{ }\frac{\partial^{n+1}\bF_c}{\partial\rms^{n+1}}(\rms_{n+1}).
\end{eqnarray}
Concatenating Eq.~(\ref{zeroth integrated partial}) with Eqs.~(\ref{general integrated partials}) leads to an expression for the retardation correction force which is the sum of a series of imbedded integrals,
\begin{eqnarray}\label{integrated force}
\bF_c(\rmt) = \bF_c^{(0)}(\rmt) +\int^{\rmt}_{\rmt_o}\rmd\rms_1\mbox{ }\frac{\rmv_{\rms_1}}{2} \frac{\rmd}{\rmd\rmR} \bF_c^{(0)}(\rms_1) +\int^{\rmt}_{\rmt_o}\rmd\rms_1\int^{\rms_1}_{\rmt_o}\rmd\rms_2\mbox{ }\frac{\rmv_{\rms_1}\rmv_{\rms_2}}{2^2} \frac{\rmd^2}{\rmd\rmR^2} \bF_c^{(0)}(\rms_2) +...\mbox{ }.
\end{eqnarray}
This result could have been written down directly since it has a straightforward interpretation of being the sum of the integrated effects of each of the partial time derivatives. Each term in Eq.~(\ref{integrated force}) can be evaluated by making a change of variables from time to position with the identity $\rmv=\rmd\rmR/\rmd\rmt$. For example, the first term gives,
\begin{eqnarray}\label{first partial integrated 1}
\int^{\rmt}_{\rmt_o}\rmd\rms_1\mbox{ }\frac{\rmv}{2} \frac{\rmd}{\rmd\rmR} \bF_c^{(0)}(\rms_1) = \int^{\rmR(\rmt)}_{\rmR(\rmt_o)}\rmd\rmR_1\mbox{ }\frac{1}{2} \frac{\rmd}{\rmd\rmR} \bF_c^{(0)}(\rmR_1)= \frac{1}{2} \bigg[\bF_c^{(0)}(\rmR_t) -\bF_c^{(0)}(\rmR_0) \bigg],
\end{eqnarray}
and further terms give,
\begin{eqnarray}\label{first partial integrated 2}
\int^{\rmt}_{\rmt_o}\rmd\rms_1\int^{\rms_1}_{\rmt_o}\rmd\rms_2 ... \int^{\rms_{n-1}}_{\rmt_o}\rmd\rms_n  \frac{\rmv_{\rms_1}\rmv_{\rms_2}..\rmv_{\rms_n}}{2^n} \mbox{ }\frac{\rmd^n}{\rmd\rmR^n}\bF_c^{(0)}(\rms_n)= \frac{1}{2^n} \bigg[\bF_c^{(0)}(\rmR_t) -\bF_c^{(0)}(\rmR_0) \bigg].
\end{eqnarray}
Substituting these into Eq.~(\ref{integrated force}) gives a geometric series with the result
\begin{eqnarray}
\label{total correction} \bF_c(\rmR) &=& \bF^{(0)}(\rmR) + \sum_{n=1}^{\infty} \bigg(\frac{1}{2}\bigg)^n \bigg( \bF^{(0)}(\rmR) -\bF^{(0)}(\rmR_0) \bigg) \nonumber\\
&=& 2\bF^{(0)}(\rmR) -\bF^{(0)}(\rmR_0),
\end{eqnarray}
where $\rmR_o=\rmR(\rmt_0)$ is the distance from the conducting wall at which the atom was originally at rest. The force $\bF^{(0)}(\rmR)$ is the stationary atom retardation correction to the vdW force.

\subsubsection{force and potential}
Inspection of Eq.~(\ref{total correction}) shows that if the atom is released but remains stationary, then the retardation force will be the stationary atom value. On the other hand if the atom is released infinitely far from the conducting wall and moves in toward the wall, then the retardation force near the wall will be twice the stationary value. At a finite initial distance the retardation force will vary between these values. The force in all cases will depend only on the position. Thus the atom still moves as if it were in a conservative potential and the potential it feels depends on where it started.

Combining the retardation correction force with the electrostatic force and simplifying as in Eq.~(\ref{stationary force}) gives the atom-wall force to be
\begin{eqnarray}
\label{adiabatic force}\bF_{am}(\rmR) = \hat{\bf e}_z \frac{\alpha_o\hbar\w_0^2}{8\pi } \bigg(\frac{\rmd}{\rmd \rmR}\bigg)^3 \frac{1}{\rmR} \int_0^\infty \frac{\rmd x}{x^2 +\w_0^2} \mbox{ }\rme^{-2\rmR x/c} -\hat{\bf e}_z \frac{\alpha_o\hbar\w_0^2}{4\pi } \bigg(\frac{\rmd}{\rmd\rmr}\bigg)^3 \int_0^\infty \frac{\rmd k}{k c+\w_0} \frac{\sin(2k\rmr)}{2k\rmr}\Bigg|^\rmR_{\rmR_0}.
\end{eqnarray}
The first term is the stationary atom-wall force and the second term is a residual force which pulls the atom back to its original point of release. The force can easily be turned into the potential which the atom feels:
\begin{eqnarray}
\label{adiabatic potential}U_{am}(\rmR) = -\frac{\alpha_o\hbar\w_0^2}{8\pi } \bigg(\frac{\rmd}{\rmd \rmR}\bigg)^2 \frac{1}{\rmR} \int_0^\infty \frac{\rmd x}{x^2 +\w_0^2} \mbox{ }\rme^{-2\rmR x/c} + \frac{\alpha_o\hbar\w_0^2}{4\pi } \bigg(\frac{\rmd}{\rmd\rmr}\bigg)^2 \int_0^\infty \frac{\rmd k}{k c+\w_0} \frac{\sin(2k\rmr)}{2k\rmr}\Bigg|^\rmR_{\rmR_0}.
\end{eqnarray}
Since the first term in the potential is the stationary atom-wall potential, in the regions near and far from the wall it will have the expected inverse powers of distance dependence, as shown in Eq.~(\ref{stationary limits}). The second term is the residual potential due to the motion.

\section{Discussion}
\subsection{Physical Interpretation}
In the energy gradient approach, one interprets the force between a polarizable atom and a wall as arising from the Lamb shift in the atomic ground state energy. Spatial variation of the ground state energy is expected to generate a force which pushes the atom to lower energy positions, but the mechanism for such a force is not given explicitly. In the final analysis, since the only players in the full system are the atom and the EMF field, such a force must come from the emission and reabsorption of photons. Our approach provides an interpretation of how a net force arises from the emission-reabsorption processes in the presence of a boundary.

The connection between the Lamb shift calculation and our calculation is the dressed ground state of the atom, which is the true ground state of the full Hamiltonian. Expanded in the free (or bare) Hamiltonian basis, the dressed ground state is a quantum superposition of bare atom-EMF states, and is often described as an atom surrounded by a cloud of virtual photons which it continually emits and reabsorbs. In the energy gradient approach, the atom-EMF is assumed to always be in the stationary dressed ground state. By contrast, in our approach a bare state is allowed to evolve quantum mechanically into the dressed ground state. The difference between these two is crucial to understanding how the coherent QED correction comes about. By allowing the atom-EMF to evolve into a dressed ground state we leave open the possibility that the motion of the atom can affect how closely to the stationary dressed ground state the system evolves. Or in the language of the virtual photon cloud, the distribution of virtually occupied modes is allowed to differ from the stationary atom case.

\subsubsection{stationary atom}
Even without motion, the atom's virtual photon cloud is altered by the presence of the wall. For a perfectly conducting wall, the TE and TM spatial mode functions of the EMF are given by Eqs.~(\ref{mode functions 1}- \ref{mode functions 2}). Those mode functions are determined by solving the wave equations with the given boundary conditions on the wall, and are constructed by linear combinations of plane wave modes. The creation and annihilation operators of the TE and TM EMF modes ($\rmb^\dagger, \rmb$) are thus combinations of the creation and annihilation operators of plane wave modes ($\rma^\dagger, \rma$) moving toward and away from the wall. Inspection of the Hamiltonian and the propagator shows that it is emission followed by absorption, which is the source of the force. In the interest of finding a physical interpretation, one can think of virtual processes in the presence of the wall in terms of plane waves. Then the emission-reabsorption of a wall-constrained mode is:
\begin{eqnarray}
\rmb_\bk \rmb_\bk^\dagger u_\bk(\bX)&\sim& (\rma_{\bk} \rme^{\rmi\bk\cdot\bX} -\rma_{-\bk} \rme^{-\rmi\bk\cdot\bX})(\rma_{\bk}^\dagger \rme^{-\rmi\bk\cdot\bX} -\rma_{-\bk}^\dagger
\rme^{\rmi\bk\cdot\bX}) \nonumber\\
&\sim& \rma_{\bk} \rma_{\bk}^\dagger +\rma_{-\bk} \rma_{-\bk}^\dagger -\rma_{-\bk} \rma_{\bk}^\dagger \rme^{-2\rmi\bk\cdot\bX} -\rma_{\bk} \rma_{-\bk}^\dagger \rme^{2\rmi\bk\cdot\bX} \nonumber
\end{eqnarray}
The first two terms are emission-reabsorption of the same photon and contribute no net momenta to the atom. The second two terms are emission of one photon and reabsorption of the reflected photon. Each of those contributes a $2\bk_z$ momentum to the atom. The effect of those processes on the force can be seen explicitly in Eq.~(\ref{general force rewrite}). The first term in Eq.~(\ref{general force rewrite}) originates from the $H_{I2}$ interaction and the second terms from the $H_{I1}$ interaction. In both terms, the sum over wavevectors is a sum over emission followed by reflected absorption processes, with each contributing a $2\bk_z$ momentum. Thus, the presence of the wall alters the atoms virtual photon cloud by reflecting some of the modes. The process of emission and reabsorption puts the photon cloud into a steady state distribution with the net effect on the atom of a retardation force.

\subsubsection{moving atom}
Once the stationary retardation force is understood in terms of the wall effect on the virtual photon cloud, the modification of it for an adiabatically moving atom can be interpreted as part of the Doppler effect. The effect is easiest to explain in the reference frame of the atom, in which it is the wall which will be moving toward or away from the atom. Then, as in the stationary case, the virtual photon cloud will be altered by reflection off the wall. However, in the case of the moving wall, the reflected photons will be Doppler shifted due to the walls motion. In the language of the virtual photon cloud, the distribution of photons around a moving atom will be Doppler shifted. This shift builds up in the photon cloud much like charge in a capacitor connected to a loop of wire in a changing magnetic field, and it can only be discharged through absorption into the atom. The net effect, over the retardation force, will be to push the atom against such built up Doppler shift, back to its original point of release.

\subsection{Prospects For Experimental Observation}
\subsubsection{Reflection From An Evanescent Laser}
A situation in which the motional modification of the retardation correction will be important is for the reflection of cold atoms off the evanescent field of an otherwise totally internally reflected laser beam. For example, in a recent experiment by Landragin et. al.~\cite{lan1}, cold alkali atoms are dropped onto a crystal with an evanescent wave running along the surface. The atom-wall interaction pulls the atoms towards the wall. The dipole potential of the evanescent wave, on the other hand, causes a repulsion of the atoms from the crystal. The combination of those two creates a barrier through which some fraction of the atoms tunnel and the rest reflect back out. The authors measure the fraction of reflected atoms versus the barrier height. As the barrier height is lowered it will at some point drop below the energy of the incoming atoms. At that point, all the atoms will be able to classically roll over the barrier, and no atoms will be reflected. The evanescent laser power required to reach that barrier height depends sensitively on the the atom-wall attraction. By comparison of measurement with theory, the authors show that the electrostatic attraction alone does not accurately predict the threshold laser power. They show that the prediction of a retardation corrected force is closer to the measured value. When we combine the motional modification to the retardation correction we are able to make a further modified prediction for the threshold. The calculations done in this paper are for a perfect conductor, not a dielectric boundary, so the modifications predicted here should not be applied directly to the case of a dielectric boundary. However, a general statement can be made that a coherent QED correction will cause a lowered prediction for the threshold laser power, since it will tend to decrease the atom-wall attraction. If one naively applies a dielectric factor to our result for the conducting plate to compensate for the difference, the present prediction for the threshold energy (14.8 $\Gamma$) is close to the measured value (14.9 $\Gamma$). Extension of the present work to a dielectric wall is ongoing.

\subsubsection{Transmission Between Parallel Plates}
Another experiment which has been able to observe the retardation of the van~der~Waals force involves a stream of ground state atoms passing between two plates~\cite{suk1}. Due to the attraction of the atoms toward the plates, some of the atoms fall onto and stick to the plates. The fraction of atoms that pass through the gap depends on the atom-wall potential. By measuring the opacity (fraction of atoms that do not pass through) for different gap widths, the authors probe the attractive atom-wall potential. This experiment holds less promise of observing a coherent QED correction to the retardation, than the previous example. The reason being that in this experiment the atoms first come into interaction with the walls at a distance of only a few resonant atomic wavelengths. The atom and EMF thus do not have as much motion over which to develop a coherent effect. Within that caveat, a general prediction can be made that the coherent correction will tend to decrease the opacity.

\subsection{Conclusion}
Our result exactly reproduces the Lamb shift result for a stationary atom. For an adiabatically slowly moving atom, a correction due to the Doppler shift is found. Our result for the retardation correction for adiabatic motion is up to twice as much as that computed for stationary atoms. Agreement with the energy gradient result in the stationary atom case shows that our non-perturbative approach captures the effects of entanglement which we sought. The physical interpretation is that the atom-EMF system evolves from an initially factorizable bare state into the interacting Hamiltonian ground state, which is an entangled state in the free Hamiltonian basis. This process is known as self-dressing. The correction for a slowly moving atom shows how our approach can go beyond Lamb shift calculations. The correction is due to the Doppler shift in that the virtual photon cloud which dresses the atom is shifted.

\appendix
\section{Recursive calculation of effective action}
The Hamiltonian is given in Eq.~(\ref{hamiltonian1}). The evaluation of the transition amplitude as a path integral begins with slicing it into infinitesimal steps. A single infinitesimal step transition amplitude for initial EMF vacuum and atomic ground state (i.e. the initial EMF and Grassmannian labels set to zero) is,
\begin{eqnarray}
\label{first infinitesmal amp} \langle \bX_1, \{\zb_{1\bk}\}, \psib_1 ;\rmt+\epsilon | \exp[-\frac{\rmi}{\hbar} H\e] | \bX_0, \{0_\bk\}, 0 ;\rmt \rangle &\mbox{}& \nonumber\\
=\exp\bigg[\frac{\rmi M(\bX_1 -\bX_0)^2\e}{2\e^2\hbar} &-& \rmi\sum_{\bk e} \psib_1 \zb_{1\bk} \frac{\g_1 \e}{\sqrt{\w_\bk}} \bp_{eg}\cdot\bu_\bk^\dagger -\rmi\sum_{\bk\bl} \zb_{1\bk} \zb_{1\bl} \frac{\la^2\e}{\sqrt{\w_\bk \w_\bl}} \bu_\bk^\dagger\cdot\bu_\bl^\dagger \bigg] \\
=\exp\bigg[\frac{\rmi M(\bX_1 -\bX_0)^2\e}{2\e^2\hbar} &+& A_1 +\sum_{\bk e} \psib_1 \zb_{1\bk} B_{1\bk e} +\sum_{\bk\bl} \zb_{1\bk} \zb_{1\bl} C_{1\bk\bl} \bigg]
\end{eqnarray}
With the obvious definitions of $A_1$, $B_{1\bk e}$, and $C_{1,\bk\bl}$. The first infinitesimal step transition amplitude, eq(\ref{first infinitesmal amp}), can be used to derive the 2 infinitesimal step amplitude:
\begin{eqnarray}
\langle \bX_2, \{\zb_{\bk 2}\}, \psib_2 ;\rmt+2\epsilon | \exp[-2\frac{\rmi}{\hbar} H\e] | \bX_0, \{0_\bk\}, 0 ;\rmt \rangle &=& \nonumber\\
\int\rmd\mu(\bX_1)\rmd\mu(\z_1)\rmd\mu(\psi_1)\langle \bX_2, \{\zb_{2\bk}\}, \psib_2 ;\rmt&+&2\epsilon | \exp[-\frac{\rmi}{\hbar} H\e]| \bX_1, \{\z_{1\bk}\}, \psi_1 ;\rmt+\epsilon \rangle  \nonumber\\
&\times&\langle \bX_1, \{\zb_{1\bk}\}, \psib_1 ;\rmt+\epsilon | \exp[-\frac{\rmi}{\hbar} H\e] | \bX_0, \{0_\bk\}, 0 ;\rmt \rangle
\end{eqnarray}
The result is:
\begin{eqnarray}
\label{2 infinitesmal amp} \langle \bX_2, \{\zb_{2\bk}\}, \psib_2 ;\rmt +2\epsilon| \exp[-2\frac{\rmi}{\hbar} H\epsilon] | \bX_0, \{0_\bk\}, 0 ;\rmt\rangle &\mbox{}& \nonumber\\
=\int\rmd\mu(\bX_1)\exp\bigg[A_2 +\sum_{\bk e} \psib_2 \zb_{2\bk} B_{2\bk e} &+&\sum_{\bk\bl} \zb_{2\bk} \zb_{2\bl} C_{2\bk\bl} +\sum_{j=1}^2 \frac{\rmi M(\bX_j -\bX_{j-1})^2\e}{2\e^2\hbar} \bigg]
\end{eqnarray}
For definitions of the coefficients see Eq.~(\ref{coefficient difference}) with $n=2$. The 2-step transition amplitude can be generalized to an n-step transition amplitude:
\begin{eqnarray}
\label{n infinitesmal amp} \langle \bX_n, \{\zb_{n\bk}\}, \psib_n ;\rmt+n\epsilon| \exp[-\frac{\rmi}{\hbar}\sum_{j=1}^{n} H_j \e] | \bX_0, \{0_\bk\}, 0 ;\rmt\rangle &\mbox{}& \nonumber\\
=\int\prod_{j=1}^n\rmd\mu(\bX_j) \exp\bigg[A_n +\sum_{\bk e} \psib_n \zb_{n\bk} B_{n\bk e} &+&\sum_{\bk\bl} \zb_{n\bk} \zb_{n\bl} C_{n\bk\bl} +\sum_{j=1}^n \frac{\rmi M(\bX_j -\bX_{j-1})^2\e}{2\e^2\hbar} \bigg]
\end{eqnarray}
with the finite difference equations:
\begin{eqnarray}\label{coefficient difference}
A_{n} &=& A_{n-1} -\rmi\e\sum_\bk \frac{\la^2}{\w_\bk}(\bu_{n\bk}^\dagger \cdot \bu_{n\bk}) -\rmi\e\sum_{\bk e} \frac{\gb_n}{\sqrt{\w_\bk}}(\bp_{ge}\cdot\bu_{n\bk}) B_{n-1,\bk e} +O(\e^2)\\
B_{n\bk e} &=& (1-\rmi\w_0\e -\rmi\w_\bk\e) \mbox{ }B_{n-1,\bk e} -\rmi\e\frac{\g_n}{\sqrt{\w_\bk}}(\bp_{eg}\cdot\bu_{n\bk}^\dagger) +\rmi\e \sum_{\bl e^\prime}  \frac{\gb_n}{\sqrt{\w_\bl}} (\bp_{ge^\prime}\cdot\bu_{n\bl}) B_{n-1,\bl e^\prime}\mbox{ }B_{n-1,\bk e} \nonumber\\
&\mbox{}&\mbox{ }\mbox{ }\mbox{ }\mbox{ }\mbox{ }\mbox{ }\mbox{ }\mbox{ }\mbox{ }\mbox{ }\mbox{ }\mbox{ }\mbox{ } -\rmi\e\sum_{\bl} \frac{2\la^2}{\sqrt{\w_\bk \w_\bl}} (\bu_{n\bk}^\dagger\cdot\bu_{n\bl}) B_{n-1,\bl e} -\rmi\e\sum_{\bl} \frac{2\g_n}{\sqrt{\w_\bl}} (\bp_{eg}\cdot\bu_{n\bl}) C_{n-1,\bk\bl} +O(\e^2)\\
C_{n\bk\bl} &=& (1 -\rmi\w_\bk\e -\rmi\w_\bl\e) C_{n-1,\bk\bl} -\rmi\e\frac{\la^2}{\sqrt{\w_\bk \w_\bl}} (\bu_{n\bk}^\dagger\cdot\bu_{n\bl}^\dagger) -\rmi\e\sum_\bq \frac{2\la^2}{\sqrt{\w_\bq \w_\bl}} (\bu_{n\bl}^\dagger\cdot\bu_{n\bq}) C_{n-1,\bk\bq} \nonumber\\
&\mbox{}&\mbox{ }\mbox{ }\mbox{ }\mbox{ }\mbox{ }\mbox{ }\mbox{ }\mbox{ }\mbox{ }\mbox{ }\mbox{ }\mbox{ }\mbox{ } -\rmi\e\sum_\bq \frac{2\la^2}{\sqrt{\w_\bq \w_\bk}} (\bu_{n\bk}^\dagger\cdot\bu_{n\bq}) C_{n-1,\bq\bl} -\rmi\e\sum_{\bq e} \frac{2\gb_n}{\sqrt{\w_\bq}} (\bp_{ge}\cdot\bu_{n\bq}) C_{n-1,\bl\bq} B_{n-1,\bk e} \nonumber\\
&\mbox{}&\mbox{ }\mbox{ }\mbox{ }\mbox{ }\mbox{ }\mbox{ }\mbox{ }\mbox{ }\mbox{ }\mbox{ }\mbox{ }\mbox{ }\mbox{ } -\rmi\e\sum_{e} \frac{\gb_n}{\sqrt{\w_\bl}} (\bp_{ge}\cdot\bu_{n\bl}^\dagger) B_{n-1,\bk e} +O(\e^2)
\end{eqnarray}
In the continuous limit those become first order differential equations with the following integral solutions:
\begin{eqnarray}
\label{differential eq a} A(\rmt+\tau) &=& -\rmi\int_{\rmt}^{\rmt+\tau} \rmd\rms\sum_\bk \frac{\la^2}{\w_\bk}(\bu_{\bk}^\dagger(\rms) \cdot \bu_{\bk}(\rms)) -\rmi\int_{\rmt}^{\rmt+\tau} \rmd\rms\sum_{\bk e} \frac{\gb}{\sqrt{\w_\bk}}(\bp_{ge}\cdot\bu_{\bk}(\rms)) B_{\bk e}(\rms)\\
\label{differential eq b} B_{\bk e}(\rmt+\tau) &=& -\rmi\int_{\rmt}^{\rmt+\tau} \rmd\rms\frac{\g}{\sqrt{\w_\bk}} \rme^{-\rmi(\w_0 +\w_\bk)(\rmt+\tau-\rms)} (\bp_{eg}\cdot\bu_{\bk}^\dagger(\rms)) +\rmi\int_{\rmt}^{\rmt+\tau} \rmd\rms\sum_{\bl e^\prime} \frac{\gb}{\sqrt{\w_\bl}} (\bp_{ge^\prime}\cdot\bu_{\bl}(\rms)) B_{\bl e^\prime}(\rms)\mbox{ }B_{\bk e}(\rms) \nonumber\\
&\mbox{}&\mbox{ }\mbox{ }\mbox{ }\mbox{ }\mbox{ }\mbox{ }\mbox{ }\mbox{ }\mbox{ }\mbox{ }\mbox{ }\mbox{ }\mbox{ } -\rmi\int_{\rmt}^{\rmt+\tau} \rmd\rms\sum_{\bl} \frac{2\la^2}{\sqrt{\w_\bk \w_\bl}} (\bu_{\bk}^\dagger(\rms)\cdot\bu_{\bl}(\rms)) B_{\bl e}(\rms) -\rmi\int_{\rmt}^{\rmt+\tau} \rmd\rms\sum_{\bl} \frac{2\g}{\sqrt{\w_\bl}} (\bp_{eg}\cdot\bu_{\bl}(\rms)) C_{\bk\bl}(\rms)\\
\label{differential eq c} C_{\bk\bl}(\rmt+\tau) &=& -\rmi\int_{\rmt}^{\rmt+\tau} \rmd\rms\frac{\la^2}{\sqrt{\w_\bk \w_\bl}} (\bu_{\bk}^\dagger(\rms)\cdot\bu_{\bl}^\dagger(\rms)) -\rmi\int_{\rmt}^{\rmt+\tau} \rmd\rms\sum_\bq \frac{2\la^2}{\sqrt{\w_\bq \w_\bl}} (\bu_{\bl}^\dagger(\rms)\cdot\bu_{\bq}(\rms)) C_{\bk\bq}(\rms) \nonumber\\
&\mbox{}&\mbox{ }\mbox{ }\mbox{ }\mbox{ }\mbox{ }\mbox{ }\mbox{ }\mbox{ }\mbox{ }\mbox{ }\mbox{ }\mbox{ }\mbox{ } -\rmi\int_{\rmt}^{\rmt+\tau} \rmd\rms\sum_\bq \frac{2\la^2}{\sqrt{\w_\bq \w_\bk}} (\bu_{\bk}^\dagger(\rms)\cdot\bu_{\bq}(\rms)) C_{\bq\bl}(\rms) -\rmi\int_{\rmt}^{\rmt+\tau} \rmd\rms\sum_{\bq e} \frac{2\gb}{\sqrt{\w_\bq}} (\bp_{ge}\cdot\bu_{\bq}(\rms))
C_{\bl\bq}(\rms) B_{\bk e}(\rms) \nonumber\\
&\mbox{}&\mbox{ }\mbox{ }\mbox{ }\mbox{ }\mbox{ }\mbox{ }\mbox{ }\mbox{ }\mbox{ }\mbox{ }\mbox{ }\mbox{ }\mbox{ } -\rmi\sum_{e} \int_{\rmt}^{\rmt+\tau} \rmd\rms\frac{\gb}{\sqrt{\w_\bl}} (\bp_{ge}\cdot\bu_{\bl}^\dagger(\rms)) B_{\bk e}(\rms)
\end{eqnarray}
The transition amplitude of Eq.~(\ref{n infinitesmal amp}) can be further simplified by setting the final EMF and atomic states to vacuum and ground, respectively. The transition amplitude is then:
\begin{eqnarray}
\label{n-step transition amp}
\langle \bX_n, \{0_\bk\}, 0 ;\rmt+\tau| \exp[-\frac{\rmi}{\hbar}\int_{\rmt}^{\rmt+\tau} H \rmd\rms] | \bX_0, \{0_\bk\}, 0 ;\rmt\rangle &=& \int D\mu(\bX(\rms)) \exp\bigg[A(\rmt+\tau) +\int_{\rmt}^{\rmt+\tau}\rmd\rms\frac{\rmi M \dot{\bX}^2(\rms)}{2\hbar} \bigg]
\end{eqnarray}
The equations for $B(\rms)$ and $C(\rms)$, Eq.~(\ref{differential eq b}) and Eq.~(\ref{differential eq c}), are Volterra type integral equations. Their solutions are infinite Born series in orders of the coupling. Approximations in the above coefficients are approximations in the basic vertex. To $O(\g^2)$:
\begin{equation}
A(\rmt+\tau) = -\rmi\int_{\rmt}^{\rmt+\tau} \rmd\rms\sum_\bk \frac{\la^2}{\w_\bk}[\bu_{\bk}^\dagger(\bX(\rms)) \cdot \bu_{\bk}(\bX(\rms))] -\int_{\rmt}^{\rmt+\tau} \rmd\rms \int_{\rmt}^{\rms} \rmd\rmr\sum_{\bk e} \frac{\g^2}{\w_\bk} \mbox{ }\rme^{-\rmi(\w_\bk +\w_0)(\rms-\rmr)}[\bu_\bk (\bX(\rms))\cdot\bp_{ge}] [\bu_\bk^*(\bX(\rmr))\cdot\bp_{eg}]
\end{equation}
The transition amplitude with an $O(\g^2)$ vertex is thus:
\begin{eqnarray}
\langle \bX_f;\rmt+\tau | \exp[-\frac{\rmi}{\hbar}\int_{\rmt}^{\rmt+\tau} H \rmd\rms] | \bX_i ;\rmt \rangle &=&\nonumber\\
\int D\bX \exp\bigg\{ \rmi\int_{\rmt}^{\rmt+\tau} \bigg[\frac{M\dot{\bX}^2}{2\hbar} &-&\sum_\bk \frac{\la^2}{\w_\bk} \bu_\bk^*(\bX(\rms))\cdot\bu_\bk(\bX(\rms)) \nonumber\\
&+&\rmi\int_{\rmt}^{\rms} \rmd\rmr\sum_{\bk e} \frac{\g^2}{\w_\bk} \mbox{ }\rme^{-\rmi(\w_\bk +\w_0)(\rms-\rmr)}[\bu_\bk (\bX(\rms))\cdot\bp_{ge}] [\bu_\bk^*(\bX(\rmr))\cdot\bp_{eg}]\bigg]\rmd\rms \bigg\}
\end{eqnarray}

In the above transition amplitude the polarization mode functions are dotted with the dipole vector of the atom. The direction that the atom's dipole vector takes will depend on the quantization direction chosen for the atom's internal state, but we are not free to choose a quantization direction. That is because the atom's dipole is induced by the vacuum fluctuations, and is free to point in any direction. In that light, choosing a particular direction seems invalid. Due to the form of the dipole - EM polarization function couplings, the induced atomic dipoles in different directions do not interfere, and a set of excited states (and thus different quantization directions) can be summed over. Such a set of independent excited states will form a resolution of unity and thus give a factor of unity contribution. The above transition amplitude can then be generalized to reflect the induced dipole:
\begin{eqnarray}
\langle \bX_f;\rmt+\tau | \exp[-\frac{\rmi}{\hbar}\int_{\rmt}^{\rmt+\tau} H \rmd\rms] | \bX_i ;\rmt \rangle &=&\nonumber\\
\int D\bX \exp\bigg\{ \rmi\int_{\rmt}^{\rmt+\tau} \bigg[\frac{M\dot{\bX}^2}{2\hbar} &-&\sum_\bk \frac{\la^2}{\w_\bk} \bu_\bk^*(\bX(\rms))\cdot\bu_\bk(\bX(\rms)) \nonumber\\
&+&\rmi\bp_z^2 \int_{\rmt}^{\rms} \rmd\rmr\sum_\bk \frac{\g^2}{\w_\bk} \mbox{ }\rme^{-\rmi(\w_\bk +\w_0)(\rms-\rmr)}\bu_\bk (\bX(\rms))\cdot\bu_\bk^*(\bX(\rmr)) +O(\rme^4) \bigg]\rmd\rms \bigg\}
\end{eqnarray}
with $\bp_z^2 = \langle {\rm g}| \bp_z^2 | {\rm g} \rangle$ (the ground state expectation value of $\bp_z^2$).

\section{Momentum Computation}
Putting in the spatial mode functions of Eq.~(\ref{mode functions 1}) into the above gives the semi-classical transition amplitude in the presence of a conducting wall.
\begin{eqnarray}
\label{classical wall amplitude}  K[\bX_f; \rmt+\tau, \bX_i; \rmt] &=& \bigg( \frac{M}{2\pi\rmi\hbar \tau}\bigg)^{3/2}\nonumber\\
\times\exp\bigg\{ &+&\frac{\rmi M(\bX_f -\bX_i)^2}{2\hbar\tau} -\frac{2\rmi\la^2}{L^3} \int_{\rmt}^{\rmt+\tau} \rmd\rms \sum_\bk \frac{1}{\w_\bk} +O(\rme^4/M) \nonumber\\
&+&\frac{\rmi\la^2}{L^3}\int_{\rmt}^{\rmt+\tau} \rmd\rms \sum_\bk \frac{\cos^2\theta}{\w_\bk} \bigg[ \rme^{2\rmi\bk_z\cdot\bX_c^0(\rms)} +\rme^{-2\rmi\bk_z\cdot\bX_c^0(\rms)} \bigg] \nonumber\\
&-& \frac{\g^2 \bp_z^2}{L^3} \int_{\rmt}^{\rmt+\tau} \rmd\rms \int_{\rmt}^{\rms} \rmd\rmr \sum_\bk \frac{1}{\w_\bk} \rme^{-\rmi(\w_\bk +\w_0)(\rms-\rmr) +\rmi\bk_{||}\cdot(\bX_c^0(\rms) -\bX_c^0(\rmr))} \bigg[ \rme^{\rmi\bk_z\cdot(\bX_c^0(\rms)-\bX_c^0(\rmr))}
+\rme^{-\rmi\bk_z\cdot(\bX_c^0(\rms)-\bX_c^0(\rmr))} \bigg] \\
&+& \frac{\g^2 \bp_z^2}{L^3} \int_{\rmt}^{\rmt+\tau} \rmd\rms \int_{\rmt}^{\rms} \rmd\rmr \sum_\bk \frac{\cos^2\theta}{\w_\bk} \rme^{-\rmi(\w_\bk +\w_0)(\rms-\rmr) +\rmi\bk_{||}\cdot(\bX_c^0(\rms) -\bX_c^0(\rmr))} \bigg[ \rme^{\rmi\bk_z\cdot(\bX_c^0(\rms)+\bX_c^0(\rmr))} +\rme^{-\rmi\bk_z\cdot(\bX_c^0(\rms)+\bX_c^0(\rmr))} \bigg]  \bigg\} \nonumber
\end{eqnarray}
With the inclusion of the conducting boundary spatial mode functions the sums over momentum space are now over the positive half space. Despite it's complicated appearance, the transition amplitude above is in a useful form for computing the evolution of the momentum expectation value. The key point is that the transition amplitude of Eq.~(\ref{classical wall amplitude}) is the product of several exponentials of exponentials, and contains only c-numbers. Therefore, each exponential can be expanded out into a series, the summands of all the series collected together, and the necessary integrations performed on the collected summand before redistributing the summand and resuming each exponential. That is, the individual exponentials in Eq.~(\ref{classical wall amplitude})
can be expanded in terms such as:
\begin{equation}
\exp\bigg\{ \frac{\rmi\la^2}{L^3}\int_{\rmt}^{\rmt+\tau} \rmd\rms \sum_\bk \frac{\cos^2\theta}{\w_\bk} \rme^{2\rmi\bk_z\cdot\bX_c^0(\rms)} \bigg\} = \bigg[ \sum_{n=0}^\infty \frac{1}{n!} \bigg(\frac{\rmi\la^2}{L^3}\bigg)^n \prod_{m=1}^n \int_{\rmt}^{\rmt+\tau} \rmd\rms_m \sum_{\bk_m} \frac{\cos^2\theta_m}{\w_{\bk_m}}\bigg] \mbox{ }\mbox{ }\mbox{ }\rme^{2\rmi\sum_m \bk_{mz}\cdot\bX_c^0(\rms_m)}
\end{equation}
\begin{eqnarray}
\exp\bigg\{ &-&\frac{\g^2 \bp_z^2}{L^3}\int_{\rmt}^{\rmt+\tau} \rmd\rms \int_{\rmt}^{\rms} \rmd\rmr \sum_\bk \frac{1}{\w_\bk} \rme^{-\rmi(\w_\bk +\w_0)(\rms-\rmr)+ \rmi\bk\cdot(\bX_c^0(\rms) -\bX_c^0(\rmr))} \bigg\}
\nonumber\\
&=& \bigg[ \sum_{n=0}^\infty \frac{1}{n!} \bigg(-\frac{\g^2 \bp_z^2}{L^3}\bigg)^n \prod_{m=1}^n \int_{\rmt}^{\rmt+\tau} \rmd\rms_m \int_{\rmt}^{\rms_m} \rmd\rmr_m \sum_{\bk_m} \frac{1}{\w_{\bk_m}} \rme^{-\rmi(\w_{\bk_m} +\w_0)(\rms_m -\rmr_m)} \bigg] \mbox{ }\mbox{ }\mbox{ }\rme^{\rmi\sum_m \bk_{m}\cdot(\bX_c^0(\rms_m) -\bX_c^0(\rmr_m))}
\end{eqnarray}
The resulting collected summand is,
\begin{eqnarray}
\mbox{Summand}(\{n\}) = \exp\bigg\{ &+&2\rmi\sum_{m_1 =1}^{n_1} \bk_{m_1 z}\cdot\bX_c^0(\rms_{m_1}) -2\rmi\sum_{m_2 =1}^{n_2} \bk_{m_2 z}\cdot\bX_c^0(\rms_{m_2}) \nonumber\\
&+&\rmi\sum_{m_3 =1}^{n_3} (\bk_{m_3 ||} +\bk_{m_3 z})\cdot(\bX_c^0(\rms_{m_3}) -\bX_c^0(\rmr_{m_3})) +\rmi\sum_{m_4 =1}^{n_4} (\bk_{m_4 ||} -\bk_{m_4 z})\cdot(\bX_c^0(\rms_{m_4}) -\bX_c^0(\rmr_{m_4})) \nonumber\\
&+&\rmi\sum_{m_5 =1}^{n_5} \bk_{m_5 ||} \cdot(\bX_c^0(\rms_{m_5}) -\bX_c^0(\rmr_{m_5})) +\rmi\sum_{m_5 =1}^{n_5} \bk_{m_5 z}\cdot(\bX_c^0(\rms_{m_5}) +\bX_c^0(\rmr_{m_5})) \nonumber\\
&+&\rmi\sum_{m_6 =1}^{n_6} \bk_{m_6 ||} \cdot(\bX_c^0(\rms_{m_6}) -\bX_c^0(\rmr_{m_6})) -\rmi\sum_{m_6 =1}^{n_6} \bk_{m_6 z} \cdot(\bX_c^0(\rms_{m_6}) +\bX_c^0(\rmr_{m_6})) \bigg\} \\
=\exp\bigg\{ \rmi{\bf c(\{n\})}\cdot(\bX_f &-&\bX_i) +\rmi{\bf b(\{n\})}\cdot\bX_i \bigg\}
\end{eqnarray}
with definitions
\begin{eqnarray}
{\bf c(\{n\})} = &+&2\sum_{m_1 =1}^{n_1} \bk_{m_1 z}\frac{\rms_{m_1}}{\tau} -2\sum_{m_2 =1}^{n_2} \bk_{m_2 z}\frac{\rms_{m_2}}{\tau} \nonumber\\
&+&\sum_{m_3 =1}^{n_3} (\bk_{m_3 ||} +\bk_{m_3 z})\frac{\rms_{m_3} -\rmr_{m_3}}{\tau} +\sum_{m_4 =1}^{n_4} (\bk_{m_4 ||} -\bk_{m_4 z})\frac{\rms_{m_4} -\rmr_{m_4}}{\tau} \nonumber\\
&+&\sum_{m_5 =1}^{n_5} \bk_{m_5 ||} \frac{\rms_{m_5} -\rmr_{m_5}}{\tau} +\sum_{m_5 =1}^{n_5} \bk_{m_5 z} \frac{\rms_{m_5} +\rmr_{m_5}}{\tau} \nonumber\\
&+&\sum_{m_6 =1}^{n_6} \bk_{m_6 ||} \frac{\rms_{m_6} -\rmr_{m_6}}{\tau} -\sum_{m_6 =1}^{n_6} \bk_{m_6 z} \frac{\rms_{m_3} +\rmr_{m_3}}{\tau}
\end{eqnarray}
and
\begin{eqnarray}
{\bf b(\{n\})} = 2\sum_{m_1 =1}^{n_1} \bk_{m_1 z} -2\sum_{m_2 =1}^{n_2} \bk_{m_2 z} +2\sum_{m_5 =1}^{n_5} \bk_{m_5 z} -2\sum_{m_6 =1}^{n_6} \bk_{m_6 z}
\end{eqnarray}
The momentum expectation value is then,
\begin{eqnarray}
\bP(\rmt+\tau) = \frac{\hbar}{N} \sum_{\{n,n^\prime\}} \Delta(\{n,n^\prime\}) \int \frac{\rmd\bP_f}{(2\pi)^3} \mbox{ }\bP_f\int\rmd\bX_i \rmd\bXp_i \rmd\bX_f \rmd\bXp_f \mbox{ }\Psi(\bX_i) \Psi^*(\bX^\prime_i) \mbox{ } \exp\bigg\{&-&\rmi\bP_f\cdot\bX_f +\rmi\bP_f\cdot\bXp_f \bigg\} \nonumber\\
\times\exp\bigg\{ \rmi{\bf c(\{n\})}\cdot(\bX_f -\bX_i) +\rmi{\bf b(\{n\})}\cdot\bX_i -\rmi{\bf c(\{n^\prime\})}\cdot(\bXp_f -\bXp_i) &-&\rmi{\bf b(\{n^\prime\})}\cdot\bXp_i \bigg\}
\end{eqnarray}
The momentum expectation value, the normalization factor, and other moments of the momentum operator can be computed with the generating function:
\begin{eqnarray}
Z(\bJ) = \sum_{\{n,n^\prime\}} \Delta(\{n,n^\prime\}) \int \frac{\rmd\bP_f}{(2\pi)^3} \int\rmd\bX_i \rmd\bXp_i \rmd\bX_f \rmd\bXp_f \mbox{ }\Psi(\bX_i) \Psi^*(\bX^\prime_i) \mbox{ } \exp\bigg\{&-&\rmi\bP_f\cdot\bX_f +\rmi\bP_f\cdot\bXp_f +\rmi\bP_f\cdot\bJ \bigg\} \nonumber\\
\times\exp\bigg\{ \rmi{\bf c(\{n\})}\cdot(\bX_f -\bX_i) +\rmi{\bf b(\{n\})}\cdot\bX_i -\rmi{\bf c(\{n^\prime\})}\cdot(\bXp_f -\bXp_i) &-&\rmi{\bf b(\{n^\prime\})}\cdot\bXp_i \bigg\}
\end{eqnarray}
from which:
\begin{equation}
\bP(\rmt+\tau) = \frac{\hbar}{\rmi Z(0)} \frac{\rmd Z(\bJ)}{\rmd\bJ} \Bigg|_{\bJ=0}
\end{equation}
The factor $\Delta(\{n,n^\prime\})$ is the summation measure. The initial wavefunction is taken to be a Gaussian centered at $(\bR, \bP_0)$ with the standard deviations $(\sigma, 1/\sigma)$. This choice allows the possibility that the atom is slowly moving toward the wall. Slowly, in this case, means adiabatically such that the external motion is much slower than internal time scales.
\begin{eqnarray}
\Psi(\bX_i) = \bigg( \frac{1}{\sqrt{\pi\sigma^2}} \bigg)^{3/2} \exp\bigg\{ -\frac{(\bX_i -\bR)^2}{2\sigma^2} +\rmi\bP_0\cdot\bX_i \bigg\}
\end{eqnarray}
In the limits $M\rightarrow\infty$ and $\sigma\rightarrow 0$ such that $\frac{\bP_0}{M}\rightarrow\bV$ and $\sigma^2
M\rightarrow\infty$ the generating function is:
\begin{eqnarray}
\label{generating function} Z(\bJ) = \exp\Bigg\{&+&\frac{\rmi\la^2}{L^3}\int_{\rmt}^{\rmt+\tau} \rmd\rms \sum_\bk \frac{\cos^2\theta}{\w_\bk} \rme^{\rmi\bk_z\cdot\bJ}\bigg[ \rme^{2\rmi\bk_z\cdot(\bR +\bV\rms)} -c.c. \bigg] \nonumber\\
&-& \frac{\g^2 \bp_z^2}{L^3} \int_{\rmt}^{\rmt+\tau} \rmd\rms \int_{\rmt}^{\rms} \rmd\rmr \sum_\bk \frac{1}{\w_\bk} \bigg[\rme^{-\rmi(\w_\bk +\w_0)(\rms-\rmr) +\rmi\bk\cdot\bV (\rms-\rmr)} +c.c. \bigg]  \\
&+& \frac{\g^2 \bp_z^2}{L^3} \int_{\rmt}^{\rmt+\tau} \rmd\rms \int_{\rmt}^{\rms} \rmd\rmr \sum_\bk \frac{\cos^2\theta}{\w_\bk} \rme^{\rmi\bk_z\cdot\bJ} \bigg[ \rme^{-\rmi(\w_\bk +\w_0)(\rms-\rmr) +\rmi\bk_z\cdot(2\bR +\bV (\rms+\rmr-2\rmt))+\rmi\bk_{||}\cdot\bV (\rms-\rmr)} +c.c. \bigg] \nonumber\\
&-&\frac{\bJ^2}{4\sigma^2} +\rmi\bJ\cdot\bP_0 +O(\rme^4/M) +O(\sigma^2) \Bigg\} \nonumber
\end{eqnarray}
Finally, in the limits $M\rightarrow\infty$ and $\sigma\rightarrow 0$ the momentum expectation value is:
\begin{eqnarray}
\bP(\rmt+\tau) = \bP_0 &-&\frac{2\rmi\la^2 \hbar}{L^3} \sum_\bk \frac{\bk_z \cos^2\theta}{\w_\bk} \int_{\rmt}^{\rmt+\tau} \rmd\rms \mbox{ }\rme^{-2\rmi\bk_z \cdot(\bR +\bV\rms)} \nonumber\\
&+&\frac{\g^2 \bp_z^2 \hbar}{L^3} \sum_\bk \frac{\bk_z \cos^2\theta}{\w_\bk} \int_{\rmt}^{\rmt+\tau} \rmd\rms \int_{\rmt}^{\rms} \rmd\rmr \mbox{ }\rme^{-\rmi\bk_z \cdot(2\bR +\bV (\rms+\rmr-2\rmt))} \bigg[ \rme^{-\rmi(\w_k +\w_0)(\rms-\rmr)} -\rme^{\rmi(\w_k +\w_0)(\rms-\rmr)} \bigg]
\end{eqnarray}

\end{document}